\documentclass{elsart}

\usepackage{graphicx}

\usepackage{amssymb}
\usepackage{amsmath}
\usepackage{amsfonts}
\usepackage{graphicx}
\usepackage{ulem} 
\usepackage[usenames]{color}

\providecommand{\av}[1]{\left\langle #1 \right\rangle} 

\begin{document}

\begin{frontmatter}



\title{Canonical equilibrium distribution derived from Helmholtz potential}

\author[label1a,label1b]{Thomas Oikonomou},
\ead{thoik@physics.uoc.gr}
\author[label2]{G. Baris Bagci} and
\ead{baris.bagci@ege.edu.tr}
\author[label2]{Ugur Tirnakli}
\ead{ugur.tirnakli@ege.edu.tr}
\address[label1a]{Department of Physics, University of Crete, 71003 Heraklion, Hellas}
\address[label1b]{Institute of Physical Chemistry, National Center for Scientific
Research ``Demokritos", 15310 Athens, Hellas}
\address[label2]{Department of Physics, Faculty of Science, Ege University, 35100
Izmir, Turkey}

\begin{abstract}
Plastino and Curado [Phys. Rev. E 72, 047103 (2005)] recently
determined the equilibrium probability distribution for the
canonical ensemble using only phenomenological thermodynamical
laws as an alternative to the entropy maximization procedure of
Jaynes.
In the current paper we present another alternative derivation of
the canonical equilibrium probability distribution, which is based
on the definition of the Helmholtz free energy (and its being
constant at the equilibrium) and the assumption of the uniqueness
of the equilibrium probability distribution. Noting that this
particular derivation is applicable for all trace-form entropies, we
also apply it to the Tsallis entropy showing that the Tsallis
entropy yields genuine inverse power laws.

\end{abstract}

\begin{keyword}

Helmholtz free energy \sep Canonical Equilibrium distribution \sep
Tsallis entropy

\PACS 05.20.-y \sep 05.70.Ln \sep 89.70.Cf

\end{keyword}

\end{frontmatter}

\section{Introduction}\label{intro}

The statistical mechanics tries to explain the thermodynamic
behavior of macroscopic systems by relying on the microscopic
constituents of the physical system. In this context, it is of
utmost importance to obtain the equilibrium distribution
associated with a particular entropy structure corresponding to
the thermodynamic properties of the system. A well-known approach
to obtain the equilibrium distribution in statistical mechanics is
the entropy maximization procedure of Jaynes (JF)
\cite{Jaynes1957}. This maximization procedure in turn stems from
the second law of thermodynamics, since entropy attains its
maximum value only at equilibrium.

On the other hand, it is necessary to be able to obtain the
equilibrium distributions purely on thermodynamic grounds by also
including some microscopic ingredients such as microstates in
order to have a comparison with the entropy maximization of
Jaynes. This necessity is twofold: First of all, it is not
straightforward to conclude the validity of second law for
generalized entropies \cite{Abe2003,Bashkirov2006}. Second, it is
apparent that even ordinary statistical second law requires a deep
and detailed study \cite{Abramo2010}.

The generalized entropies do not change the inherent dynamics so
that the first law of thermodynamics (FLT) and its solutions must
be certainly preserved i.e., they must be form invariant
independent of the entropy structure one adopts \cite{Mendes}.
This is also required, since the first law of thermodynamics is
the conservation of energy \cite{Callen}. The second law of
thermodynamics too must be preserved in these generalized schemes
in order to construct a link between information-theoretic entropy
measures and the statistical mechanics. In short, then, the
generalized entropy measures must conform to the thermodynamical
laws in order to be considered useful in statistical mechanics,
instead of being solely useful for information theoretical
approaches.

The paper is organized as follows.
The next section outlines how to obtain canonical equilibrium
probability distributions through the requirement of a constant
Helmholtz potential $F$ and the assumption of the uniqueness of
the aforementioned distribution. This approach is then extended to
all trace-form entropies and applied to the Tsallis entropy
measure \cite{Tsnewbook}. Conclusions are presented in the last
section.

\section{Thermodynamics and Equilibrium Distributions}\label{sec:FLT1}
%
The first law of thermodynamics for reversible processes reads
\begin{equation}\label{FLT-1}
dU=T\,dS- P\,dV\,+ \mu\, dN\,,
\end{equation}
where $U$ is the internal energy, $S$ is the entropy and $T$ is
the temperature, while $P$ and $V$ denote pressure and volume,
respectively. The term $\mu$ is the chemical potential associated
with the change in the number of particles $N$.
A particular case of great interest in thermodynamics is when
$U=U(S,V,N)$ is a homogeneous function of first degree
\cite{Callen}. In this case, Eq.~(\ref{FLT-1}) can be explicitly
solved, independent from the analytical expressions of $T,\,P$ and
$\mu$ \cite{comment1}, yielding
\begin{equation}\label{step2}
U=TS-PV+\mu N\,.
\end{equation}
The equation above is referred to as the Euler equation of
thermodynamics. In terms of the free energy $F=F(S,V,N)$ (also
called the Helmholtz potential), the above equation can simply be
rewritten as
\begin{equation}\label{step3}
F=U-TS\,.
\end{equation}
Keeping in mind that $F=-PV+\mu N$, Eq.~(\ref{step3}) is
equivalent to Eq.~(\ref{step2}). This definition of Helmholtz
potential $F$ is also in accordance with the rest of the
thermodynamical potentials. For example, the Gibbs potential $G$
is equal to $F+PV$ so that one obtains the well known relation
$G=\mu N$. We can further rewrite Eq.~(\ref{step3}) as
\begin{equation}\label{exactFLT}
S=\beta U-\beta F\,.
\end{equation}
where $1/\beta\equiv T$.

Up to this point, all these equations contain nothing related to
the statistical nature of the physical system under investigation.
In order to introduce in Eq.~(\ref{exactFLT}) statistics, one
needs some statistical ingredients. We then consider the very
definition of Boltzmann-Gibbs (BG) statistical entropy, which
reads (setting $k_{\mathrm{B}}=1$)
\begin{equation}
S_{\mathrm{BG}}=\sum_ip_i\ln(1/p_i)\,,
\end{equation}
and the definition of the internal energy in terms of the
microscopic probabilities i.e., $U=\sum_{i}p_{i}\varepsilon_{i}$,
where $p_i$ is the probability of the corresponding $i$th
microstate and $\varepsilon_{i}$ is the microstate energies.

Then, when the Helmholtz potential is constant, Eq.
(\ref{exactFLT}) can be rewritten as
\begin{equation}\label{exactFLT1}
\sum_ip_i\ln(1/p_i)=\beta \sum_{i}p_{i}\varepsilon_{i}-\beta
\sum_{i}p_{i}F,
\end{equation}
where we also assumed probability normalization as usual.
Eq.~(\ref{exactFLT1}) naturally yields to the relation below
\begin{equation}\label{exactFLT2}
\sum\limits_{i}p_{i}\Big[ \ln \left( 1/p_{i}\right) -\beta
\varepsilon _{i}+\beta F\Big] = 0 .
\end{equation}
However, it is not trivial to obtain the equilibrium probability
distribution $p^\star_{i}$ from the equation above, since whole
summation might be equal to zero, and not only one singled out
term within the summation.
On the other hand, assuming the existence of the thermodynamic
state of the stationary equilibrium, we can invoke the uniqueness
of the aforementioned state and therefore of the corresponding
equilibrium distribution. Due to the former assumption i.e., the
existence of the equilibrium state, we have $p^\star_{i}\neq0$,
since the condition $p^\star_{i}=0$ contradicts with the existence
of an equilibrium state.
The uniqueness of the thermodynamic state of equilibrium and hence
of the equilibrium distribution implies the trivial solution to
Eq.~(\ref{exactFLT2}), equating the term in the brackets to zero
i.e.,
\begin{equation}\label{exactFLT3}
\ln \left( 1/p^\star_{i}\right) -\beta \varepsilon_{i}+\beta F =0.
\end{equation}
It is then obvious that the equation above yields an equilibrium
distribution and this is all one needs, since the equilibrium
distribution (and the state of equilibrium) is unique, so that
having \textit{one} solution of Eq.~(\ref{exactFLT2}) is
tantamount to obtaining the  \textit{only} solution indeed.
Therefore, we finally have the following equilibrium distribution
\begin{equation}\label{exactFLT4}
p^\star_{i}=\exp \big( -\beta \left( \varepsilon _{i}-F\right)
\big),
\end{equation}
which is the canonical equilibrium distribution. Moreover, note
that identifying $e^{-\beta F}$ as the canonical partition
function $Z$, one obtains another well known relation i.e.,
$F=-T\ln Z$ (remember that we set Boltzmann constant
$k_{\mathrm{B}}$ to unity).
We observe that the canonical distribution $p_i^\star$ satisfies
Eq. (\ref{exactFLT}) and the FLT in Eq. (\ref{FLT-1}).

Some remarks are in order: although mathematically simple, the
above calculations are novel in the sense that it does not make
use of ordinary entropy maximization procedure put forth by
Jaynes.
Apart from the definitions of entropy and internal energy in terms
of microscopic constituents, it relies solely on the definition of
Helmholtz free energy and the existence and the uniqueness of the
thermodynamic equilibrium state.
Moreover, obtained in this manner, one sees that the canonical
BG-equilibrium distribution, apart from the multiplicative factor
$\beta$, is of exponential form, where the exponent is formed by
shifting the microstate energies $\varepsilon_{i}$ by the constant
(and minimum at the canonical equilibrium state) Helmholtz free
energy $F$.

The above calculation can easily be extended to generalized
entropy measures, since these definitions are
information-theoretic and not supposed to violate any of the
ingredients used in the above derivation. In order to extend the
formalism above to generalized trace-form entropies, we study the
following statistical representation of the thermodynamic entropy
i.e., $S=\av{\Lambda(1/p_i)}_{\mathrm{lin}}$, where
$\av{(\cdots)_i}_\mathrm{lin} = \sum_ip_i(\cdots)_i$ is the
arithmetic expectation value of the quantity $(\cdots)_i$, $p_i$
are the microstate probabilities, and assuming that the internal
energy is given statistically as
$U=\av{\varepsilon_i}_{\mathrm{lin}}$, where $\varepsilon_i$ are
the microstate  energies. Similar to Eq.~(\ref{exactFLT1}), we now
write
%
\begin{equation}\label{StatRepres}
\av{\Lambda(1/p_i)}_{\mathrm{lin}}=\av{\beta\varepsilon_i}_{\mathrm{lin}}-\av{\beta
F}_{\mathrm{lin}} \qquad\Longrightarrow\qquad
\av{\Lambda(1/p_i)-\beta\varepsilon_i+\beta F}_{\mathrm{lin}}=0\,.
\end{equation}
%
Note that $\Lambda(1/p_i)$ is a generic function, which can be any
monotonic increasing function compatible with the linear averaging
scheme, be it ordinary or generalized logarithm.
>From the relation above, we can obtain the equilibrium probability
distribution $p_{i}$, which is compatible with FLT, namely
%
\begin{subequations}
\begin{align}\label{MaxProb}
p^\star_{i}&=\frac{1}{\Lambda^{-1}\big(\beta(\varepsilon_i-F)\big)}
 =\frac{[\Lambda^{-1}(\beta\varepsilon_i)]^{-1}}{Z}\,,\\
\label{MaxProb-b}
S&=\beta(U-F)=\sum_ip^\star_{i}[\beta\varepsilon_i\oplus_\Lambda\Lambda(Z)]\,,
\end{align}
\end{subequations}
%
where $x\oplus_\Lambda y:=\Lambda(\Lambda^{-1}(x)\Lambda^{-1}(y))$
($x,y\geq0$) and $\oplus_{\Lambda\rightarrow\ln}\rightarrow+$ is
the ordinary addition operation.
$Z$ in Eq.~(\ref{MaxProb}) represents the partition function,
defined implicitly through Eq.~(\ref{MaxProb-b}).
As can be seen, its analytical expression depends on the function
set $\{\Lambda, \Lambda^{-1}\}$.
If we now substitute the probability distribution given by
Eq.~(\ref{MaxProb}) into the left hand side of
Eq.~(\ref{exactFLT}), it reproduces indeed the Euler equation.

\subsection{Tsallis entropy}\label{subsec:Tsallis}
An entropy measure of current interest is the Tsallis entropy
\cite{Tsnewbook,Bagci1}.
This entropy can be constructed by choosing
\begin{equation}
\Lambda(q,x)\equiv\ln_q(x)=\frac{x^{1-q}-1}{1-q}\,,\quad
\Lambda^{-1}(q,x)\equiv\exp_q(x)=\big[1+(1-q)x\big]_+^{\frac{1}{1-q}}
\end{equation}
where $[x]_+=\mathrm{max}\{0,x\}$, so that the structure $S$ tends
to the following $q$-entropy
\begin{equation}
S_q=\sum_ip_i\ln_q(1/p_i)=\frac{\sum_ip_i^q-1}{1-q}\,.
\end{equation}
The equilibrium probability distribution, according to Eqs.
(\ref{MaxProb}) and (\ref{MaxProb-b}), takes the form
\begin{equation}\label{Ts-Prob-a}
\frac{1}{p^\star_{i}}=\exp_q\big(\beta(\varepsilon_i-F)\big)=Z_q\,\exp_q(\beta\varepsilon_i)\,,
\qquad\mathrm{with}\qquad Z_q=\exp_q\bigg(-\frac{\beta
F}{1+(1-q)\beta U}\bigg)\,.
\end{equation}
We note that this equilibrium distribution structure is exactly
the one obtained through the entropy maximization procedure of the
Tsallis entropy with linear constraints \cite{Tsnewbook,Bagci2}.
It is also worth noting that this method apparently does not apply
to non-trace form entropies.

\section{Conclusions}\label{Concl}

In this work, we derived the well-known canonical equilibrium
probability distribution by using the microscopic definitions of
entropy and internal energy together with the definition of the
Helmholtz free energy (and its being constant at the equilibrium)
and the existence/uniqueness of the state of equilibrium.
Providing a generalization of this derivation to all trace-form
entropies, we also obtained the stationary equilibrium probability
distribution of the Tsallis entropy which is found to be an
inverse power law distribution. This work can be considered as an
alternative to both the entropy maximization procedure of Jaynes
\cite{Jaynes1957} and recently proposed procedure of Plastino
\textit{et al}. \cite{Curado,Curado2,Curado3,Curado4,Curado5}
based on the phenomenological laws of thermodynamics.

\section*{Acknowledgments}
%
We thank an anonymous reviewer for his/her very explanatory
remarks. T.O. acknowledges partial support by the THALES Project
MACOMSYS, funded by the ESPA Program of the Ministry of Education
of Hellas.


\begin{thebibliography}{0}
\bibitem{comment1} Note that in this case the functions
$T=T(S,V,N),\,P=P(S,V,N)$ and $\mu=\mu(S,V,N)$ are homogeneous of
degree zero, as a consequence of the degree of homogeneity of
$U=U(S,V,N)$ being one.
\end{thebibliography}
\end{document}